\documentclass[fleqn,twoside]{elsart3}
\usepackage{graphicx}
\usepackage[figuresright]{rotating}

\newcommand{\AmS}{{\protect\the\textfont2
  A\kern-.1667em\lower.5ex\hbox{M}\kern-.125emS}}
\newcommand{\pl}{$\frame{+}$}

\begin{document}

\begin{frontmatter}
\title{Half-metallic Zinc-Blende Compounds}
\author[IFF]{Ph.~Mavropoulos}
\ead{Ph.Mavropoulos@fz-juelich.de}
\author[IFF]{I.~Galanakis}
\author[IFF]{P.~H.~Dederichs}
\address[IFF]{ IFF, Forschungszentrum J\"ulich, D-52425 J\"ulich,
  Germany}
\begin{abstract}
  We report first-principles calculations of zinc-blende half-metals,
  identifying systems for epitaxial growth with semiconductors, and
  present calculations for CrAs/GaAs multilayers. We find that
  half-metallicity is conserved troughout the heterostructure, making
  this a good candidate for spintronics applications.
\end{abstract}
\end{frontmatter}


Half-metallicity is the property of some spin polarised materials to
exhibit a metallic density of states (DOS) for one spin direction but
a band gap around the Fermi level $E_F$ for the other. Recently,
ordered zinc-blende (z-b) CrAs and CrSb have been fabricated by
molecular beam epitaxy~\cite{Akinaga00,Mizuguchi02b}. For these
systems experiment and calculations suggest half-metallicity. The high
Curie temperature (over 400 K) makes these compounds attractive also
for applications.  CrAs/GaAs multilayers have been also
realised~\cite{Mizuguchi02b}, showing that coherent heterostructures
are possible.
Such z-b compounds of transition elements with group-V and VI
atoms have been also studied by {\it ab-initio}
methods~\cite{Sanvito00,Galanakis03}.  

In a recent
paper~\cite{Galanakis03} we examined possible combinations of V, Cr,
and Mn, with group-V and VI elements, and found that some of them are
half-metallic at their equilibrium lattice constants, which also fit
reasonably well to those of some semiconductors (SC) as presented in
Table~\ref{Table:1}.
\begin{table}
\begin{center}
\caption{\label{Table:1} Calculated properties of
  z-b compounds at the experimental lattice constants (in parentheses)
  of z-b III-V semiconductors. A ``$+$'' means that a system is
  half-metallic, a ``$-$'' that $E_F$ is above the gap and a ``$\pm$''
  that $E_F$ is only slightly in the conduction band. The calculated
  equilibrium lattice parameter a$_{\mathrm{eq}}$ is also given. The cases when the
  lattice mismatch is small (accounting also for LDA overbinding of
  about 2\%) are marked by squares.}
\begin{tabular}{lccccccccc}
\hline
Compound  & VAs  & VSb  & CrP  & CrAs & CrSb & MnP  & MnAs & MnSb \\
a$_{\mathrm{eq}}$(\AA )   &5.54  & 5.98 & 5.19 & 5.52 & 5.92 & 5.00 & 5.36 & 5.88 \\
\hline
GaAs(5.65)& \pl  & $-$  & $+$  & \pl  & $-$  & $-$  & $-$  & $-$   \\
InP(5.87) & $+$  & $+$  & $+$  & $+$  & $+$  & $+$  & $+$  & $-$   \\
InAs(6.06)& $+$  & \pl  & $+$  & $+$  & \pl  & $+$  & $+$  & $-$   \\
GaSb(6.10)& $+$  & \pl  & $+$  & $+$  & \pl  & $+$  & $+$  & \frame{$\pm$} \\
\hline
\end{tabular}
\end{center}
\end{table}
Here we examine the case of CrAs/GaAs (001) multilayers, since the
lattice mismatch should be small. We assume that the structure has the
experimental GaAs lattice constant (5.65 \AA) and that the z-b
structure is kept throughout. The system consists of 4 monolayers (ML)
of CrAs followed by 4 ML of GaAs and periodically repeated, in
accordance with the experimental result that only a few ML can exist
within the periodically repeated supercell~\cite{Mizuguchi02b}. In the
$[$001$]$ direction of growth, this corresponds to interchanging
monoatomic layers: ...Cr/As/Cr/As/Ga/As/Ga/As... . For the
calculations we use the KKR Green function method, within the local
density approximation (LDA) of density-functional theory. This is
known to underestimate the SC band gap, even more if relativistic
effects are taken into account. For this reason we chose to make our
calculations non-relativistic; otherwise the details are as described
in Ref.~\cite{Galanakis03}.

\begin{figure}[t]
  \begin{center}
    \leavevmode
    \includegraphics[angle=90,width=7.5cm]{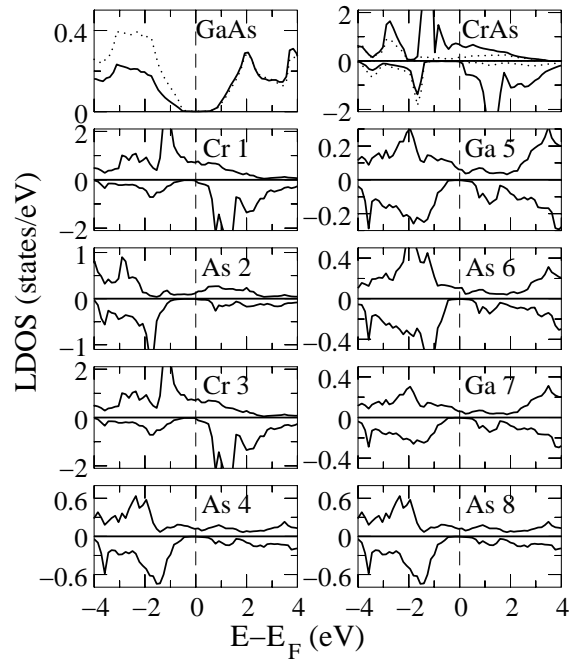}
    \caption{Atom-resolved DOS of a CrAs/GaAs
      (001) multilayer in supercell geometry. The inset numbers refer
      to the enumeration of successive layers. In the top pannels the
      bulk DOS of GaAs and CrAs is shown (Cr and Ga, with full lines,
      As with dotted). The following atoms have equivalent environment
      and DOS: Cr1 and Cr3, As4 and As8, Ga5 and Ga7. Negative numbers
      in the DOS refer to minority spin.}
    \label{fig:1}
  \end{center}
\end{figure}

Our results are presented in Figure~\ref{fig:1}. As explained in
Ref.~\cite{Galanakis03}, in bulk CrAs there is a hybridisation and
bonding-antibonding splitting of the As $p$ states with the Cr $d$
states of the $t_{2g}$ subspace ({\it i.e.} $d_{xy}$, $d_{yz}$, and
$d_{xz}$). The bonding and antibonding hybrides form wide bands, with
a clear splitting in energy. In-between, the $d$ states of the $e_g$
subspace ($d_{z^2}$ and $d_{x^2-y^2}$) form narrower bands and partly
occupy the bonding-antibonding gap. Due to the strong exchange
splitting, the occupation of the bands strongly differs for the two
spin directions: the bonding $p$-$d$ hybrides are occupied for both
spins, but the $e_g$ and partly the antibonding $p$-$d$ hybrides are
occupied only for majority spin, while the minority $e_g$ and
antibonding $p$-$d$ bands stay unoccupied. Thus the gap is formed
around $E_F$, only for minority electrons, between the bonding $p$-$d$
and the $e_g$ band.

In the multilayer the minority-spin gap is conserved throughout; no
interface states are formed for the minority spin.  This can be
understood since the growth is coherent, so that the local environment
of the interface Cr atom is not changed in CrAs/GaAs. On the other
hand, even for bulk CrAs $E_F$ is very close to the conduction band
(of $e_g$ character) for minority spin.  As a result, half-metallicity
is practically preserved throughout the multilayer, with $E_F$ merely
touching the conduction band edge (seen in the Cr1 and Cr3 layers).
Note that, even if $E_F$ would slightly enter into the minority
conduction band, these states are very localised due to their $e_g$
character, so that they would not be important for spin-dependent
transport.  For the majority spin the local DOS at $E_F$ decays within
the SC layers. For a thick SC spacer it should vanish far from the
interface, since the SC band gap is present and the DOS comes from
exponentially decaying metal-induced gap states.

\begin{thebibliography}{99}

\bibitem{Akinaga00}
H. Akinaga {\it et al.}, Jpn. J. Appl. Phys. \textbf{39}, L1118
(2000);
J. H. Zhao {\it et al.}, Appl. Phys. Lett. \textbf{79}, 2776 (2001);
K. Ono {\it et al.}, J. Appl. Phys. {\bf 91}, 8088 (2002);
M. Mizuguchi {\it et al.}, J. Appl. Phys. \textbf{91},
7917 (2002).


\bibitem{Mizuguchi02b}
M. Mizuguchi {\it et al.},
J. Magn. Magn. Mater. {\bf 239}, 269 (2002).




\bibitem{Sanvito00}
S. Sanvito and N. A. Hill, Phys. Rev. B {\bf 62}, 15553 (2000);
M. Shirai, Physica E \textbf{10}, 143 (2001);
A. Continenza {\it et al.}, Phys. Rev. B {\bf 64}, 085204 (2001);
A. Continenza {\it et al.}, Phys. Rev. B {\bf 64}, 085204 (2001);
Y.-J. Zhao {\it et al.}, Phys. Rev. B {\bf 65}, 113202 (2002);
I. Galanakis, Phys. Rev. B  \textbf{66}, 012406 (2002);
Y.-G. Xu {\it et al.}, Phys. Rev. B {\bf 66}, 184435 (2002).







\bibitem{Galanakis03}
I. Galanakis and Ph. Mavropoulos, Phys. Rev. B {\bf 67}, 104417 (2003).




\end{thebibliography}
\end{document}